\documentclass[a4paper,11pt,fleqn]{article}

\usepackage[latin1]{inputenc}
\usepackage[english]{babel}
\usepackage{amsmath,amssymb}
\usepackage{epsfig,graphicx}

\newcommand{\beq}{\begin{equation}}
\newcommand{\eeq}{\end{equation}}
\newcommand{\bea}{\begin{eqnarray}}
\newcommand{\eea}{\end{eqnarray}}
\newcommand{\bdm}{\begin{displaymath}}
\newcommand{\edm}{\end{displaymath}}
\newcommand{\eps}{\varepsilon}

\newcommand{\kt}{\textrm{\textbf{k}}}

\def \d {{\rm d }}

\newcommand{\lp}{\left(}
\newcommand{\rp}{\right)}

\def\as{\alpha_s}

\def\m{{\cal M}}

\def\ord{{\cal O}}
\def\eps{\varepsilon}

\def \msb{\overline{\textrm{MS}}}

\def\kt {\bold{k}}
\def\eps {\varepsilon}

\begin{document}
\begin{flushright}
IFUM-970-FT\\
MAN/HEP/2011/02
\end{flushright}

\begin{center}
{\Large \bf
Finite fermion mass effects in pseudoscalar Higgs production via gluon-gluon fusion}
\vspace*{1.5cm}

Fabrizio Caola$^{a}$ and Simone Marzani$^b$
\\
\vspace{0.3cm}  {\it
{}$^a$Dipartimento di Fisica, Universit\`a di Milano and
INFN, Sezione di Milano,\\
Via Celoria 16, I-20133 Milano, Italy\\ \medskip
{}$^b$School of Physics \& Astronomy, University of Manchester,\\
Oxford Road, Manchester, M13 9PL, England, U.K.}\\
\vspace*{1.5cm}

\bigskip
\bigskip

{\bf \large Abstract:}
\end{center}
We compute the leading logarithmic behaviour of the cross-section  for the production of a pseudoscalar Higgs boson in gluon-gluon fusion to all-orders in perturbation theory, in the limit of large partonic centre of mass energy. We also calculate the Higgs rapidity distribution to the same accuracy. We include the contributions of top and bottom quarks, together with their interference. Our results are given in terms of single and double integrals, evaluated explicitly up to next-to next-to leading order (NNLO). 
We use our results to improve  the known NNLO  inclusive cross-section computed in the effective theory where the fermions in the loop are integrated out.  The size of finite fermion mass effects on the inclusive cross-section is found to be small, reaching a few percent only for large values of the pseudoscalar mass.

\section{Introduction}
The first collisions at the Large Hadron Collider (LHC) at CERN marked the beginning of a new era in particle physics. 
The four primary experiments at the LHC have started taking data at energies never explored before. 
Their main target is to understand electro-weak symmetry breaking and the origin of mass in the universe. Over forty years after its proposal, the Higgs mechanism for giving mass to fundamental particles will finally be tested. 
Supersymmetric theories generally predict a richer Higgs sector than the Standard Model (SM). In the Minimal Supersymmetric Standard Model (MSSM) for instance one introduces two complex Higgs doublets, which originate five physical Higgs bosons: two neutral scalars ($h$, $H$), two charged scalars ($H^{\pm}$) and, finally, a neutral CP-odd state ($A$). 

The most important production channel for the pseudoscalar Higgs boson at hadron colliders is gluon-gluon fusion, via a fermion loop, and associated production with jets. However, the phenomenology is much richer than the case of the SM Higgs, because the coupling of the pseudoscalar to the up-type quarks decreases with $\tan \beta = v_1/v_2$ (the ratio of the two vacuum expectation values), while the one to down-type quarks increases. As a consequence the coupling to the $b$-quark becomes important and eventually dominates the top contribution for $\tan \beta \ge 10$ (for a recent review see for instance Ref.~\cite{LHCHiggs}).

In collinear factorization, the production cross-section for $gg \to A$ can be written as
\beq
\sigma(\as; x_h,m_A;m_t,m_b) = \sum_{i,j=q, \bar{q},g} \int \frac{d x_1}{x_1} \int \frac{d x_2}{x_2} \sigma_{ij}\left(\as; \frac{x_h}{x_1 x_2}, m_A ; m_t,m_b\right)f_i(x_1,m_A) f_j(x_2,m_A),
\eeq
where $x_h={m_A^2}/{s}$ and both renormalization and factorization scales have been set equal to the pseudoscalar mass $m_A$.  The partonic cross-section $\sigma_{ij}$ has been computed in QCD at next-to-leading order (NLO)~\cite{SpiraNLO}:
\beq
\sigma_{ij}\left(\as; x, m_A ; m_t,m_b\right) = \sum_{f,f'= t,b}\sigma_0^A(m_A;m_f,m_{f'}) \left[ \delta_{ij} \delta(1-x) + \frac{\as}{\pi} C^{(1)}_{ij}\left(x;\frac{m_f}{m_A},\frac{m_{f'}}{m_A}\right)+ \ord(\as^2)\right]
\eeq
The NLO corrections turned out to be large ($\ord(70 \%)$) and motivated a NNLO calculation. However, a full treatment would require the evaluation of $2 \to 1$ diagrams at three loops with massive lines. For this reason the NNLO corrections have been computed in an effective theory~\cite{EFTpseudo} in which one integrates out the contribution of the heavy fermion running in the loop~\cite{HarlanderPseudo, AnastasiouPseudo,RavindranNNLO}. This approximation at NLO works very well, far beyond its naive range of validity, essentially because the cross-section is dominated by the emission of relatively soft gluons which cannot resolve the fermion loop.  However, in the MSSM this approximation is more delicate because the mass of the pseudoscalar can be much larger than the top mass. Furthermore, at large $\tan \beta$, the coupling to the $b$-quark becomes increasingly important and the use of the effective theory less justified.

It is therefore important to quantitatively asses the role of finite fermion mass corrections beyond NLO.  In the case of SM model Higgs production, recent analyses~\cite{harlander, pak, SimoneHarlander} confirm that also at NNLO finite fermion mass effects are negligible. Those analyses are based on an asymptotic expansion of the Feynman diagrams in inverse powers of the heavy fermion mass $m_f$. However, it was soon realised~\cite{dawsonexp} that this expansion fails to converge in the limit of large partonic centre of mass energy $\hat{s}$ (i.e. small $x=m^2_{\rm Higgs}/\hat{s}$), and it gives rise to a spurious power growth. Fortunately the small-$x$ bevahiour of the partonic coefficient functions can be computed using a technique called $k_T$-factorization~\cite{CataniHQ, CataniDIS}. This has been done in the case of the SM Higgs in~\cite{SimoneHiggs,SimoneHiggsProc} and combined than with the asymptotic expansion in Refs.~\cite{harlander, pak,SimoneHarlander}.

In this letter we explicitly compute the small-$x$ behaviour of the partonic cross-section for $gg\to A$ at NNLO QCD and we match it to the effective theory result, which can be viewed as the zeroth order in the asymptotic expansion. This enables us to give a first estimate of finite fermion mass effects in NNLO pseudoscalar production. 

\section{The off-shell cross-section}
We use $k_T$-factorization~\cite{CataniHQ, CataniDIS} to determine the leading logarithmic behaviour at small-$x$ (LL$x$) of the  partonic coefficient functions to any desired order in the strong coupling. We start by considering gluon-initiated subprocesses only, as the other partonic channels can be derived using colour-charge relations.
Thus,  we compute the leading order cross-section for the production of a pseudoscalar Higgs, keeping the incoming gluons off their mass-shell
\beq
g^*(k_1,\mu,A,\epsilon_1) + g^* (k_2,\nu,B,\epsilon_2) \longrightarrow A(q),
\eeq
where $A,B$ are colour indices and $\epsilon_i$ are the polarization vectors, which will be specified later on. The leading order amplitude is obtained  by computing the one-loop massive three-point function
\beq
\mathcal{M}= \sum_f  \left[\mathcal{A}_f(k_1,k_2) + \mathcal{A}_f(k_2,k_1) \right],
\eeq
where the index $f$ runs over the fermionic flavours in the loop. In our study we consider $f=t,b$.
The coupling of the pseudoscalar Higgs is proportional to the mass of the fermion running in the loop: $-i \frac{m_f g_A}{v_0}\gamma_5$, where $v_0=\sqrt{v_1^2+v_2^2}$ and $g_{Af}$ is a dimensionless quantity, which in the MSSM, for instance, is  $g_{At}= 1/\tan \beta$ for up-type quarks and $g_{Ab}= \tan \beta$ for down-type ones.
The calculation of the amplitude is then straightforward;  in $D=4-2 \eps$ we have
\bea
\mathcal{A}_f &=& (-i g_s \mu^{2\eps})^2 \frac{-i m_f g_{Af}}{v_0} 
\text{Tr}\left[
T_A T_B
\right]
i^3 \times \nonumber\\
&\times&
\int \frac{d^D l}{(2\pi)^D}
\frac{
\text{Tr}\left[ (\not l + m_f)\not \epsilon_1 
(\not l - \not k_1 + m_f) \gamma_5
(\not l + \not k_2 + m_f) \not \epsilon_2
 \right]}
{\lp l^2-m_f^2 \rp \lp (l-k_1)^2-m_f^2\rp \lp (l+k_2)^2-m_f^2\rp}.
\eea
The treatment of $\gamma_5$ and of the Levi-Civita tensor away from four dimensions is in principle a delicate problem~\cite{thooft, maison, axialanomaly}. 
This computation is analogous to calculation of the axial anomaly. It is well known that, although the result of the loop integral is finite,  there are some issues regarding the regularization procedure one must take care of.
With this in mind, using standard manipulation of Dirac matrices and Feynman parametrization of the loop integrand, we obtain:  
\bea \label{offshellampl}
\mathcal{A}_f&=& -4i m_f g_{Af} \epsilon(k_1,k_2,\epsilon_1, \epsilon_2) 
(g_s \mu^{2\eps})^2 \frac{m_f}{v_0} 
\lp T_R \delta^{AB} \rp \times \nonumber \\
&\times&
\int \frac{d^D l}{(2\pi)^D} d\alpha d\beta
\frac{\Gamma(3)}
{
\left[l^2-\lp m_f^2 - \alpha(1-\alpha) k_1^2 - \beta(1-\beta) k_2^2 
-2 \alpha \beta k_1\cdot k_2 \rp \right]^3
}=\nonumber\\
&=& -4i \epsilon(k_1,k_2,\epsilon_1, \epsilon_2) 
(g_s \mu^{2\eps})^2 \frac{m_f^2 g_{Af}}{v_0} 
\lp T_R \delta^{AB} \rp \times 
i \frac{\pi^{D/2}}{(2\pi)^D} r_\Gamma \mu^{-2\epsilon} 
I_3^D(k_1^2,k_2^2,q^2, m_f^2, m_f^2, m_f^2)
\eea
where
\beq
\epsilon(k_1,k_2,\epsilon_1, \epsilon_2)= \epsilon_{\mu \nu \rho \sigma}k_1^{\mu} k_2^{\nu}\epsilon_1^{\rho} \epsilon_2^{\sigma} \quad {\rm and} \quad   r_\Gamma \equiv \frac {\Gamma^2(1-\eps) \Gamma(1+\eps)}
{\Gamma(1-2\eps)}.
\eeq
We notice that this result has a simpler form than the one for the production of a scalar Higgs (see for instance Ref.\cite{SimoneHiggs}) and it is basically given by the one-loop three-point function $I_3^D$~\cite{ellisqcd}. Since everything is regular, we can safely go back to $D=4$ dimension and we obtain
\beq
\mathcal{M}=\sum_{f=t,b} 2 \mathcal{A}_f = \sum_{f=t,b} \delta^{AB}
\frac{\as g_{Af}}{\pi} \frac{m_f^2}{v_0}
\epsilon(k_1,k_2,\epsilon_1, \epsilon_2)
I_3(k_1^2,k_2^2,q^2, m_f^2, m_f^2, m_f^2).
\eeq
In order to determine the high-energy behaviour of the cross-section it is convenient to write the kinematics in terms of:
\bea
k_i &=& z_i p_i +  \kt_i, \quad
p_i^2 =0,  \quad
k_i^2 =- |\kt_i|^2, \quad i=1,2 ,\nonumber \\ 
m_A^2 &=&q^2=(k_1+k_2)^2, \quad \hat{s} =2 p_1 \cdot p_2;
\eea
we also find useful to introduce the following dimensionless ratios:
\beq \label{dimensionlessratios}
x = \frac{m_A^2}{\hat{s}}\,, \quad \tau_f=\frac{4 m_f^2}{m_A^2}\, ,\quad \xi_i = \frac{|\kt_i|^2}{m_A^2}, \; i=1,2.
\eeq
In this notation the one-loop amplitude looks like
\beq \label{me_off}
\m= \sum_{f=b,t} \delta^{AB} \frac{\as g_A}{v_0}  \epsilon(k_1,k_2,\epsilon_1,\epsilon_2) 4 \pi \tau_f C_0 (\xi_1,\xi_2;\tau_f),
\eeq
where to make contact with the case of a scalar Higgs boson of Ref.\cite{SimoneHiggs}, we have introduced the dimensionless form factor
\beq
C_0(\xi_1,\xi_2;\tau_f)= \frac{m_f^2}{4 \pi^2 \tau_f} I_3(k_1^2,k_2^2,q^2, m_f^2, m_f^2, m_f^2).
\eeq

In order to obtain the off-shell cross section in $k_T$-factorization, we specify the polarization vectors $\epsilon_{i}$ to be
\beq \label{polarization}
\epsilon_i =  \sqrt{2}\frac{k_{i,T}^{\mu}}{k_{i,T}^2}, \quad i=1,2;
\eeq
we then square the matrix element Eq.(\ref{me_off}), sum (average) over final (initial) state colour and spin and multiply by the phase space factor $d \mathcal{P}$, obtaining
\bea \label{xsec_off}
\sigma(x, \xi_1,\xi_2;\tau_t, \tau_b) &=& \frac{1}{2 \hat{s}} \int d \mathcal{P}\overline{ | \m |^2} \nonumber \\
&=& 2 \pi^3\as^2 G_F \sqrt{2}  \int_0^{2 \pi} \frac{d \varphi}{2 \pi} \Big \{ {g_{At}}^2 \tau_t^2|C_0(\xi_1,\xi_2;\tau_t)|^2+g_{Ab}^2 \tau_b^2|C_0(\xi_1,\xi_2;\tau_b)|^2
\nonumber \\ &&+ 2 g_{At} g_{Ab}\tau_t \tau_b {\rm Re}\, C_0^*(\xi_1,\xi_2;\tau_t)C_0(\xi_1,\xi_2;\tau_b)  \Big\}
 \frac{\sin^2 \varphi}{4x} \nonumber \\ &&\times \delta \left(\frac{1}{x}-1-\xi_1-\xi_2-2 \sqrt{\xi_1 \xi_2} \cos \varphi \right), \nonumber\\
\eea
where we have introduced the Fermi constant $G_F=\frac{1}{\sqrt{2} v_0^2}$ and the azimuthal angle $\varphi$ between the transverse components of the incoming momenta.
The small-$x$ leading logarithms of the collinearly factorized coefficient function can be computed by considering an impact factor, which is defined as the triple Mellin transform of the off-shell cross-section Eq.~(\ref{xsec_off}):
\bea \label{impactfull}
&&h(N,M_1,M_2; \tau_t, \tau_b) = M_1 M_2 \int_0^1 d x x^{N-1} \int_0^{\infty} d \xi_1 \xi_1^{M_1-1} \int_0^{\infty} d \xi_2 \xi_2^{M_2-1} \sigma(x, \xi_1,\xi_2;\tau_t, \tau_b) \nonumber \\
&&=\int_0^{\infty} d \xi_1 \xi_1^{M_1-1} \int_0^{\infty} d \xi_2 \xi_2^{M_2-1} 
 \frac{\pi^3}{2}\as^2 G_F \sqrt{2}  
  \Big \{ {g_{At}}^2 \tau_t^2|C_0(\xi_1,\xi_2;\tau_t)|^2+g_{Ab}^2 \tau_b^2|C_0(\xi_1,\xi_2;\tau_b)|^2
\nonumber \\ &&+ 2 g_{At}g_{Ab}\tau_t \tau_b {\rm Re}\, C_0^*(\xi_1,\xi_2;\tau_t)C_0(\xi_1,\xi_2;\tau_b)  \Big\}
 \int_0^{2 \pi} \frac{d \varphi}{2 \pi} \frac{\sin^2 \varphi}{\left(1+\xi_1+\xi_2+2 \sqrt{\xi_1 \xi_2} \cos \varphi  \right)^N},
\eea
where we have performed the $N$ Mellin using the delta function. We remind the reader that the above impact factor leads to a result for the collinear factorized coefficient functions which is not in $\msb$. The mismatching between the two factorization schemes is known and described by an overall factor $R$~\cite{CataniDIS}. However, we are interested in the expansion of the resummed result at NNLO and the scheme dependence starts one order higher:
\beq
R= 1+ \ord(\as^3).
\eeq
Therefore, at the accuracy we are working at, we can safely ignore all issues concerning the scheme dependence of our results.

\section{Determination of the high-energy behaviour}
The leading small-$x$ singularities of the coefficient function can be now determined from the impact factor $h(N,M_1,M_2; \tau_t, \tau_b)$, identifying $M_1=M_2=\gamma_s(\as/N)$, the anomalous dimension which is dual to the leading order BFKL kernel~\cite{duality}:
\beq \label{duality}
\chi_0\left( \gamma_s \left(\frac{\as}{N}\right) \right)= \frac{N}{\as}, \quad \gamma_s\left(\frac{\as}{N}\right)= \frac{\as C_A}{\pi N}+ \ord\left( \frac{\as^4}{N^4}\right).
\eeq
and then expanding in powers of $\as$ (i.e. in powers of $M_1$, $M_2$), taking the $N\to 0$ limit. Before computing the $M$ Mellin transforms  in the full theory, it is instructive to look at the result in the heavy fermion approximation. In this effective theory we expect a spurious double logarithmic behaviour at small-$x$~\cite{CataniHQ, HautmannHiggs}. We investigate this case in the following section.
\subsection{Heavy fermion approximation}
We consider the case where there is only one fermionic flavour running in the loop, which is infinitely massive.  We have
\beq \label{limit}
\lim_{\tau_f \to \infty} C_0(\xi_1,\xi_2;\tau_f) = - \frac{1}{8 \pi^2 \tau_f} + \ord \left(\frac{1}{\tau_f^2} \right),
\eeq
so the impact factor becomes
\bea
h(N,M_1,M_2;\infty)&=&\frac{g_{Af}^2 \alpha_s^2 G_F \sqrt{2}}{128 \pi}  \int_0^{\infty} d \xi_1 \xi_1^{M_1-1} \int_0^{\infty} d \xi_2 \xi_2^{M_2-1}\frac{1}{(1+\xi_1+\xi_2)^N}
\nonumber \\ &&\int_0^{2 \pi} \frac{d \varphi}{2 \pi} \frac{\sin^2 \varphi}{\left(1+ \frac{2\sqrt{\xi_1 \xi_2}}{1+\xi_1+\xi_2}\cos \varphi\right)^N  }. 
\eea
From the expression above it is clear that one cannot set $N=0$, because the $\xi_i$ integrals would then diverge for every value of $M_1$, $M_2$.  The situation is totally analogous to scalar Higgs production and double high-energy logarithms appear to all orders in perturbation theory~\cite{HautmannHiggs}. The angular integration can be performed in terms of hypergeometric functions, but it is easy to see that it is actually safe to set $N=0$ inside the angular integral. The azimuthal integration then trivially gives:
\beq \label{all_EFT}
h(N,M_1,M_2;\infty)= \sigma_0^A \int_0^{\infty} d \xi_1 \xi_1^{M_1-1} \int_0^{\infty} d \xi_2 \xi_2^{M_2-1}
 \frac{1}{(1+\xi_1+\xi_2)^N} \left[ 1+ \ord(N)\right],
\eeq
where we have introduced the leading order cross-section in the heavy fermion limit
\beq \label{LO_EFT}
\sigma_0^A = \frac{g_{Af}^2 \alpha_s^2 G_F \sqrt{2}}{256 \pi} .
\eeq
We note that the pseudoscalar cross-section differs from the scalar one by a numerical factor $9/4$. This is expected as the two effective Lagrangians are
\bea
\mathcal{L}_{\rm eff}^H &=& \frac{\as}{12 \pi} \sqrt{G_F \sqrt{2}} G_{\mu \nu}^a G^{a \mu \nu} H, \nonumber\\
\mathcal{L}_{\rm eff}^A &=& \frac{\as g_A}{8 \pi} \sqrt{G_F \sqrt{2}} G_{\mu \nu}^a \widetilde{G}^{a \mu \nu} A.
\eea
In the heavy fermion approximation, once we have factored out the leading order result, the high-energy leading (double) logarithms  are the same for scalar and pseudoscalar Higgs:
\bea
 h(N,M_1,M_2; \infty) & = &\sigma_0^A  \left[1+\frac{M_1+M_2}{N}+\left(\frac{M_1+M_2}{N}\right)^2+\dots \right] \nonumber\\
 &=&\sigma_0^A \left[1+\frac{\as}{\pi}\frac{2 C_A}{N^2}+\left(\frac{\as}{\pi}\right)^2
 \left(\frac{2 C_A}{N^2}\right)^2+\dots \right].
 \eea
 The LL$x$ result in $x$-space is easily obtained by inverting the Mellin transform:
 \beq \label{gg_EFT}
 \sigma_{gg}(x;\infty)=  \sigma_0^A  \left[ \delta(1-x) + \frac{\as}{\pi} 2 C_A \ln \frac{1}{x} +\left(\frac{\as}{\pi} \right)^2 \frac{4 C_A^2}{6 ! }\ln^3 \frac{1}{x} + \ord(\as^3) \right] .
  \eeq
 So far we have been considering the gluon-gluon channel only. The generalization to all the other partonic subprocesses is straightforward in the high-energy limit~\cite{CataniHQ, CataniDIS}. We obtain
 \bea \label{other_EFT}
 \sigma_{qg}(x;\infty) &=&  \sigma_0^A  \left[  \frac{\as}{\pi}  C_F \ln \frac{1}{x} +\left(\frac{\as}{\pi} \right)^2 \frac{3 C_A C_F}{6 ! }\ln^3 \frac{1}{x} + \ord(\as^3) \right] , \nonumber \\
 \sigma_{qq}(x;\infty) &=&  \sigma_0^A  \left[  \left(\frac{\as}{\pi} \right)^2 \frac{2 C_F^2}{6 ! }\ln^3 \frac{1}{x} + \ord(\as^3) \right] .
  \eea
  The above results Eq.~(\ref{gg_EFT}),~(\ref{other_EFT}) agree with the small-$x$ limit of the known NNLO coefficient functions~\cite{AnastasiouPseudo}.
 
\subsection{Finite fermion masses}
We now go back to Eq.~(\ref{impactfull}) in order to compute the LL$x$ behaviour in the full theory, with top and bottom quarks running in the loop. We first notice that the presence of the form factor $|C_0|^2$ ensures that the $M$ Mellin transforms have finite radius of convergence when $N=0$.  
We write the impact factor as the sum of different contributions: the ones coming from the square of each diagram and then interference ones:
\beq \label{dec}
h(0,M_1,M_2;\tau_t,\tau_b) =\sum_{f,f'=t,b} \tilde{h}(0,M_1,M_2;\tau_f, \tau_{f'}) .
\eeq
where we have introduced:
\bea
&&\tilde{h}(0,M_1,M_2;\tau_f, \tau_{f'})= \quad\quad\quad\quad\quad\quad\quad\quad\quad\quad\quad\quad\quad\quad\quad  \nonumber \\
&&=   \frac{\pi^3\as^2 G_F g_{Af} g_{Af'}\tau_f \tau_{f'} }{2\sqrt{2} \left( 1+ \delta_{ff'}\right)}M_1 M_2 \int_0^{\infty} d \xi_1 \xi_1^{M_1-1} \int_0^{\infty} d \xi_2 \xi_2^{M_2-1}   C_0^*(\xi_1,\xi_2;\tau_f) C_0(\xi_1,\xi_2;\tau_{f'}) \nonumber \\
&&=    \frac{\sigma_0 ^A(\tau_f,\tau_{f'})}{1+ \delta_{ff'}} \Bigg[  1 - (M_1+M_2) \int_0^{\infty}  d \xi_1 \ln \xi_1 \frac{ {\rm d} }{{\rm d} \xi_1}  \frac{C_0^*(\xi_1,0;\tau_f) C_0(\xi_1,0;\tau_{f'})}{C_0^*(0,0;\tau_f) C_0(0,0;\tau_{f'})}  \nonumber \\ &&
- \frac{1}{2}(M_1^2+M_2^2) \int_0^{\infty}  d \xi_1 \ln^2 \xi_1 \frac{C_0^*(\xi_1,0;\tau_f) C_0(\xi_1,0;\tau_{f'})}{C_0^*(0,0;\tau_f) C_0(0,0;\tau_{f'})}\nonumber \\ &&
 + M_1M_2 \int_0^{\infty} d \xi_1 \int_0^{\infty} d \xi_2 \ln \xi_1 \ln \xi_2 \frac{\partial^2}{\partial \xi_1 \partial \xi_2}  \frac{C_0^*(\xi_1,\xi_2;\tau_f) C_0(\xi_1,\xi_2;\tau_{f'})}{C_0^*(0,0;\tau_f) C_0(0,0;\tau_{f'})}+ \dots \Bigg] .
\eea
with
\beq
\sigma_0^A(\tau_f,\tau_{f'}) =   \frac{\pi^3g_{At} g_{Ab}\as^2 G_F   \tau_f \tau_{f'}}{2\sqrt{2}} C_0^*(0,0;\tau_f) C_0(0,0;\tau_{f'}).
\eeq
In particular when $\tau_f= \tau_{f'}$, the LO contribution to the cross-section is
\beq
\sigma_0^A(\tau_f,\tau_f) =   \frac{\pi^3g_{Af}^2\as^2 G_F   \tau_f^2}{2\sqrt{2}} |C_0(0,0;\tau_f)|^2=  \frac{g_{Af}^2 \as^2 \sqrt{2}}{256 \pi}  |\tau_f f(\tau_f)|^2,
\eeq
\beq\label{szerodef}
 f(\tau_f)=\left\{\begin{matrix}&-\frac{1}{4}\left[\ln \left(\frac{1+\sqrt{1-\tau_f}}{1-\sqrt{1-\tau_f}}\right)- i\pi\right]^2\, & \textrm{if} \quad \tau_f < 1 \\
& \, \arcsin^2 \left(\sqrt{\frac{1}{\tau_f}}\,\right) &\textrm{if} \quad
\tau_f \ge 1.\end{matrix}\right.
\eeq

The high-energy behaviour is then determined by the identification $M_1=M_2= \gamma_s\left(\as/N \right)$. The result in momentum space is then easily recovered by inversion of the $N$-Mellin transform. We obtain
 \bea \label{gg_full}
\sigma_{gg}(x;\tau_t,\tau_b)&=&\sum_{f,f'=t,b}\frac{\sigma_0 ^A(\tau_f,\tau_{f'})}{1+ \delta_{ff'}} \Bigg[ \delta(1-x) + \frac{\as}{\pi} 2 C_A c_1(\tau_f,\tau_{f'})
\nonumber \\  && +\left(\frac{\as}{\pi} \right)^2 C_A^2 \left(2 c_2(\tau_f,\tau_{f'})+c_{11}(\tau_{f},\tau_{f'})\right) \ln \frac{1}{x}    + \ord(\as^3) \Bigg],
  \eea
  where the coefficients are expressed in terms of the following integrals
  \bea \label{numcoeffNLO}
  c_1(\tau_f, \tau_{f'}) &=& - \int_0^{\infty}  d \xi_1 \ln \xi_1 \frac{ {\rm d} }{{\rm d} \xi_1}  \frac{C_0^*(\xi_1,0;\tau_f) C_0(\xi_1,0;\tau_{f'})}{C_0^*(0,0;\tau_f) C_0(0,0;\tau_{f'})},
  \eea
  \bea  \label{numcoeffNNLO}
  c_2(\tau_f, \tau_{f'}) &=& - \frac{1}{2}\int_0^{\infty}  d \xi_1 \ln^2 \xi_1 \frac{ {\rm d} }{{\rm d} \xi_1} \frac{C_0^*(\xi_1,0;\tau_f) C_0(\xi_1,0;\tau_{f'})}{C_0^*(0,0;\tau_f) C_0(0,0;\tau_{f'})} , \nonumber \\
  c_{11}(\tau_f, \tau_{f'}) &=& \int_0^{\infty} d \xi_1 \int_0^{\infty} d \xi_2 \ln \xi_1 \ln \xi_2 \frac{\partial^2}{\partial \xi_1 \partial \xi_2} \frac{C_0^*(\xi_1,\xi_2;\tau_f) C_0(\xi_1,\xi_2;\tau_{f'})}{C_0^*(0,0;\tau_f) C_0(0,0;\tau_{f'})}.
  \eea
  We evaluate numerically the above integrals; the results are tabulated in Table~\ref{table1} and \ref{table2}. We use $m_b= 4.19$~GeV and $m_t= 172.0$~GeV.
  The LL$x$ behaviour of the other subprocesses is easily obtained with colour-charge relations:
 \bea\label{other_full}
 \sigma_{qg}(x,\tau_t,\tau_b)&=&  
\sum_{f,f'=t,b}\frac{\sigma_0 ^A(\tau_f,\tau_{f'})}{1+ \delta_{ff'}}  \Bigg[\frac{\as}{\pi}  C_F c_1(\tau_f,\tau_{f'})
 +\left(\frac{\as}{\pi} \right)^2 C_A C_F \left(c_2(\tau_f,\tau_{f'})+c_{11}(\tau_{f},\tau_{f'})\right) \ln \frac{1}{x}   \nonumber \\ && + \ord(\as^3) \Bigg], \nonumber\\
 \sigma_{q_iq_j(\bar{q}_j)}(x,\tau_t,\tau_b)&=&  
\sum_{f,f'=t,b}\frac{\sigma_0 ^A(\tau_f,\tau_{f'})}{1+ \delta_{ff'}} \Bigg[
 +\left(\frac{\as}{\pi} \right)^2 C_F^2 \left(c_{11}(\tau_{f},\tau_{f'})\right) \ln \frac{1}{x}    + \ord(\as^3) \Bigg].
   \eea
  
\begin{figure}
\begin{center}
\includegraphics[width=0.5\textwidth,angle=-90,clip]{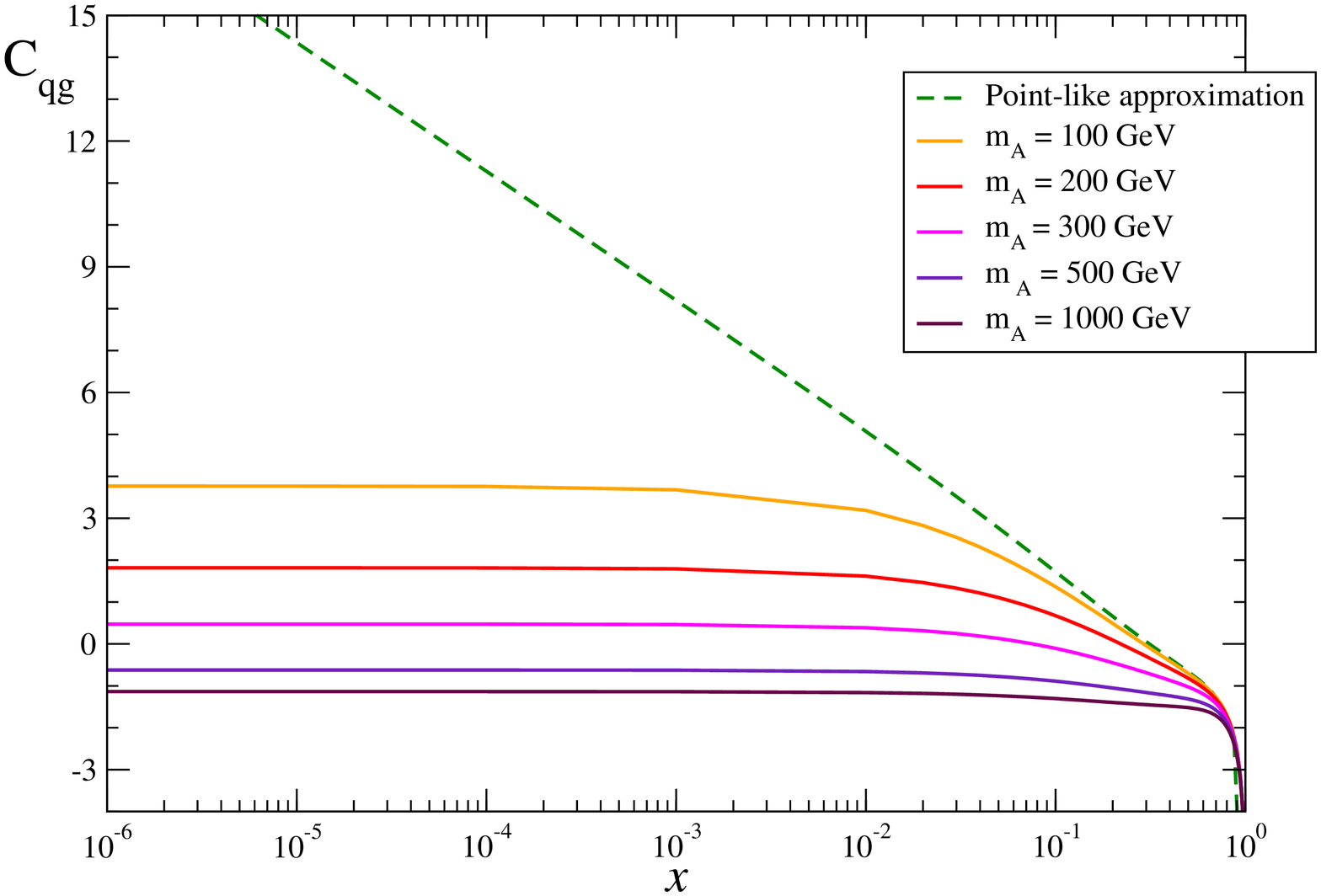}
\caption{The NLO coefficient function in the quark-gluon channel $C^{(1)}_{qg}$ in the case of just a top quark running in the loop, as a function of $x$. The green dashed curve is the effective theory result, while the solid curves are full result~\cite{SpiraNLO} for $m_t=172.0$~GeV and $m_A= 100, 200,300,500$ and $1000$~GeV (from top to bottom)}\label{NLO}\end{center}
\end{figure}

  We note that when the fermion mass is kept finite, the cross-section for pseudoscalar Higgs production has the expected single-logarithmic behaviour.  The coefficients of the LL$x$ singularities differ from the scalar Higgs case~\cite{SimoneHarlander}.  
  As a check of our procedure we have compared our asymptotic behaviour to the small-$x$ limit of the full NLO calculation of Ref.~\cite{SpiraNLO} and we have found perfect agreement.  As an example in Fig.~\ref{NLO} we plot the $\ord(\as)$ coefficient function in the quark-gluon channel $C^{(1)}_{qg}$ in the case of just a top quark running in the loop, as a function of $x$. The green dashed curve is the effective theory result, which exhibits the spurious logarithmic growth at small $x$, while the solid curves are for $m_t=172.0$~GeV and $m_A= 100, 200,300,500$~and~$1000$~GeV respectively. When the fermion mass in the loop is kept finite the coefficient function has the expected constant behaviour at small $x$, which is in numerical agreement with the coefficients in Table~\ref{table1}, once multiplied by the appropriate colour factor $C_F$.
\begin{table} 
\centering
\begin{tabular}{|c||c|c|c|}
\hline
$m_A$ & $c_1(\tau_b,\tau_b)$ & $c_1(\tau_t,\tau_t)$ & $c_1(\tau_t,\tau_b)+c_1(\tau_b,\tau_t)$ \\\hline
\hline
100   & -1.9831& 2.8251&  -2.7106        \\\hline
200   & -2.6344 &   1.3629& -3.8880      \\\hline
300    &-3.0289 &   0.35479 &    -4.6358 \\\hline
400    & -3.3123 & -0.37258 &   -5.0118     \\\hline
500     & -3.5334 &-0.46678 &   -5.2424    \\\hline
600     & -3.7147 &-0.55464 &  -5.4629   \\\hline
700        &  -3.8681 &-0.63658&     -5.6627       \\\hline
800       & -4.0014 & -0.71311 &   -5.8438       \\\hline
900       &  -4.1190 &-0.78479 &-6.0089      \\\hline
1000        & -4.2244 &  -0.85211 &  -6.1605 \\\hline
\end{tabular}
\caption{Results for the NLO coefficient $c_1$, obtained by numerical evaluation of the integrals in Eq.~(\ref{numcoeffNLO}).} \label{table1}
\end{table}

\begin{table}
\centering
\begin{tabular}{|c||c|c|c||c|c|c|}
\hline
$m_A$ & $c_2(\tau_b,\tau_b)$ & $c_2(\tau_t,\tau_t)$ & $c_2(\tau_t,\tau_b)+c_2(\tau_b,\tau_t)$ &
 $c_{11}(\tau_b,\tau_b)$ & $c_{11}(\tau_t,\tau_t)$ & $c_{11}(\tau_t,\tau_b)+c_{11}(\tau_b,\tau_t)$ \\\hline
\hline
100   &4.1527 & 5.5035 &  5.3683 &  2.1777 & 8.5243 &  -0.9321         \\\hline
200   & 6.2393 &  2.4237 &  8.0566     &  4.7678&  2.4575 &1.4378  \\\hline
300    & 7.7494 & 1.5146 &   10.1403     &  6.7586 & 0.78853 &  4.0076 \\\hline
400    &8.9476 & 1.4378& 11.8815    &  8.3770 & 0.70446&  5.5309 \\\hline
500     & 9.9485 &  1.4748 &    13.1381     &  9.7473 &  0.47723 &  6.1455\\\hline
600     &  10.8118 & 1.5279 &  14.2203   &  10.9403 & 0.32497 &   6.8722  \\\hline
700        & 11.5735 & 1.5921&     15.1778     &  11.9994& 0.22632&  7.6113  \\\hline
800       &  12.2566 & 1.6637 &     16.0402     &  12.9539 &  0.16702 &  8.3396  \\\hline
900       &  12.8771& 1.7404 & 16.8276      &  13.8239 & 0.13720 &  9.0487    \\\hline
1000        & 13.4462 & 1.8205 & 17.5537    & 14.6245 & 0.12990 &  9.7366\\\hline
\end{tabular}
\caption{Results  for the NNLO coefficients $c_2$ and
$c_{11}$, obtained by numerical evaluation of the integrals in Eq.~(\ref{numcoeffNNLO}).}  \label{table2}
\end{table}

\subsection{Rapidity distribution}
The formalism which enables one to resum high-energy enhanced contributions has been recently extended to the case of rapidity distributions~\cite{smallxrap}. The production of a pseudoscalar Higgs via gluon gluon fusion is particularly simple because the leading order process is $2\to 1$ and thus the kinematics is trivial. The case of scalar Higgs production has been studied in~\cite{smallxrap}; the structure of the result is the same for pseudoscalar case, the only difference being the values of the numerical coefficients computed in Tables~\ref{table1} and~\ref{table2}. 
We write the rapidity distribution at the parton level as
\beq
\frac{\d\sigma_{ab}}{\d y}(x,y;\tau_t,\tau_b) = \sum_{f,f'=t,b} \frac{\d\tilde{\sigma}_{ab}}{\d y}(x,y,\tau_f,\tau_{f'}).
\eeq
The small-$x$ limit of the different contributions are
\bea
\frac{\d\tilde{\sigma}_{gg}}{\d y}(x,y;\tau_f,\tau_{f'}) &=&   \frac{\sigma_0 ^A(\tau_f,\tau_{f'})}{1+ \delta_{ff'}}  \Bigg \{  \frac{1}{2}\delta(1-x) \left[\delta
\lp y - \frac 1 2 \ln x \rp + \delta
\lp y + \frac 1 2 \ln x \rp
\right] \nonumber \\ &&+
 \frac{\as}{\pi}  C_A \left[\delta
\lp y - \frac 1 2 \ln x \rp + \delta
\lp y + \frac 1 2 \ln x \rp
\right]
c_1(\tau_f,\tau_{f'}) \nonumber \\ &&+
 \left(\frac{\as}{\pi}\right)^2  C_A^2\Bigg[
\left[ \delta
\lp y - \frac 1 2 \ln x \rp +\delta
\lp y + \frac 1 2 \ln x \rp
\right]  \ln \frac{1}{x} c_{2}(\tau_f,\tau_{f'}) \nonumber \\ &&
+c_{11}(\tau_f,\tau_{f'}) \Bigg]  + \ord(\as^3) \Bigg \},  \nonumber \\
\frac{\d\tilde{\sigma}_{qg}}{\d y}(x,y;\tau_f,\tau_{f'}) &=&\frac{\sigma_0 ^A(\tau_f,\tau_{f'})}{1+ \delta_{ff'}} 
\Bigg \{  \frac{\as}{\pi}  C_F \left[ \delta
\lp y - \frac 1 2 \ln x \rp + \delta
\lp y + \frac 1 2 \ln x \rp
\right]
\frac{c_1(\tau_f,\tau_{f'})}{2} \nonumber \\  &&+  \left(\frac{\as}{\pi}\right)^2  C_A C_F \Bigg[
\left[ \delta
\lp y - \frac 1 2 \ln x \rp + \delta
\lp y + \frac 1 2 \ln x \rp
\right]  \ln \frac{1}{x} \frac{c_{2}(\tau_f,\tau_{f'})}{2} \nonumber \\ &&+c_{11}(\tau) \Bigg]  + \ord(\as^3) \Bigg \}, \nonumber \\
\frac{\d\tilde{\sigma}_{q_iq_j(\bar q_j)}}{\d y}(x,y;\tau_f,\tau_{f'}) &=& \frac{\sigma_0 ^A(\tau_f,\tau_{f'})}{1+ \delta_{ff'}} 
\Bigg \{ \left(\frac{\as}{\pi}\right)^2  C_F^2  c_{11}(\tau_f,\tau_{f'}) + \ord(\as^3)
\Bigg \}.
\eea

\section{Improvement of the NLO and NNLO cross-sections}
In the spirit of Ref.~\cite{SimoneHiggs,SimoneHiggsProc} we construct an approximated cross-section by matching the coefficient functions computed in the effective theory $C^{(n)}_{ij}(x; \infty)$, which are valid at large $x$, to the asymptotic behaviour at small $x$, computed keeping the mass of the fermions finite $C^{(n)}_{ij,{\rm small}\, x}(x;\tau_f)$. We concentrate first on the case of a top loop, which gives the dominant contribution at small $\tan \beta$. We define
\beq \label{approx}
C^{(n)}_{ij}(x; \tau_t)=C^{(n)}_{ij}(x; \infty)+  \Theta(x_0-x) \left[C^{(n)}_{ij,{\rm small}\, x}(x;\tau_t)- \lim_{x\to0} C^{(n)}_{ij}(x; \infty) \right] \,, \quad n=1,2.
\eeq 

We start by considering the NLO case $(n=1)$, where the full result is already known.  The spurious logarithmic growth of the effective theory at small $x$ must be replaced by a constant behaviour. The matching point $x_0$ can be chosen at the intersection of the $m_f\to \infty$ curve with the asymptotic constant. 
Subleading terms, which vanish at small-$x$, can be adjusted to ensure continuity. 

\begin{figure}
\begin{center}
\includegraphics[width=0.5\textwidth,angle=-90,clip]{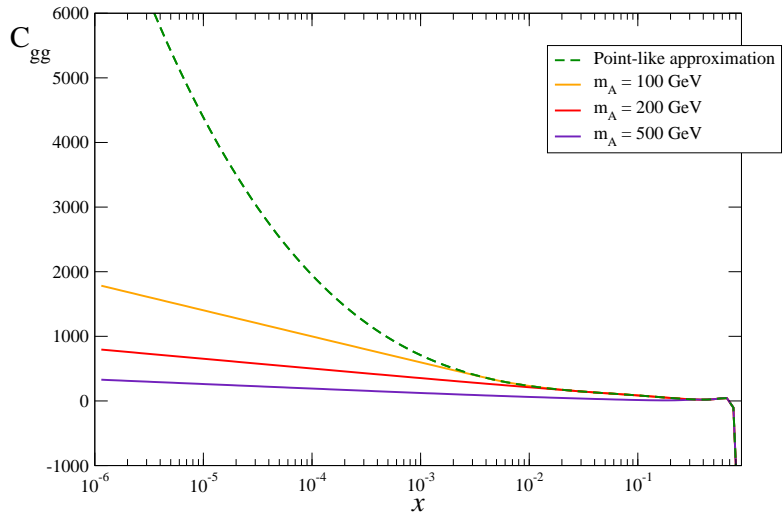}
\caption{The NNLO coefficient function in the gluon-gluon channel $C^{(2)}_{gg}$ in the case of just a top quark running in the loop, as a function of $x$. The green dashed curve is the effective theory result, while the solid curves are our matched results Eq.~(\ref{approx}) with $m_t=172.0$~GeV and $m_A= 100, 200,300$ and $500$~GeV (from top to bottom).}\label{coeff_func_NNLO}\end{center}
\end{figure}

At the NNLO level $(n=2)$ the 
effective theory exhibits a cubic logarithmic behaviour, while the full theory has only a single logarithm. The approximated cross-section is constructed subtracting from the effective theory result all those terms which do not vanish at small $x$.
However, at this order, the matching is more ambiguous because we only control the LL$x$ behaviour, i.e. the coefficient of the single logarithm but we do not know the constant. Following~\cite{SimoneHiggsProc} we choose to match the two curves at the point where the slope of the effective theory curve, on a logarithmic scale, approaches the correct asymptotic behaviour. The constant is then adjusted requiring continuity of the coefficient function. This method ensures that the approximated coefficient functions Eq.~(\ref{approx}) and their first derivatives are continuous functions of $x$. As an example, in Fig.~\ref{coeff_func_NNLO} we plot the NNLO coefficient function for the gluon-gluon subprocess, in the effective theory and in our approximation for different values of $m_A=100,200$~and~$500$~GeV, in case of a top quark running in the loop. 
In case of small ${\rm tan} \beta$ we can neglect the contribution coming from $b$-quark loops.
In order to asses the numerical impact of finite top mass effect coming from the hard part of the partonic coefficient functions, we compute $K$-factors, defined as
\beq \label{kfactdef}
K^{\rm t } (\alpha_s; x_h, m_A;m_t) = \frac{\sigma^{\rm matched}(\alpha_s; x_h, m_A;m_t)}{\sigma(\alpha_s; x_h, m_A;\infty)},
\eeq
using MSTW08 NNLO parton distribution functions~\cite{mstw08} for proton-proton collisions at $\sqrt{s}= 14$~TeV. We obtain:
\bea
K^{\rm t } &=& 1.000, \quad m_A= 100\, {\rm GeV}, \nonumber \\
K^{\rm t }  &=& 0.998, \quad m_A= 200\, {\rm GeV} ,\nonumber \\
K^{\rm t }  &=& 0.985, \quad m_A= 500\, {\rm GeV},
\eea
The $K$-factors clearly shows that at  small $\tan \beta$, where the top quark contribution dominates, finite fermion mass effects are small in a wide range of the pseudoscalar mass, reaching a few percent only above the two-top threshold.

Our calculation of the LL$x$ behaviour also includes the effect of a bottom quark running in the loop as well as the top-bottom interference. These contributions cannot be neglected at large $\tan \beta$. However, at first sight, it is not clear how to use this information in a NNLO matched calculation. The mass of the $b$ quark is much smaller than $m_A$ and hence an effective theory where this contribution is integrated out is not justified. However, we notice that the large $x$ behaviour of the coefficient function, being dictated by soft gluon effects, does not depend on the fermion mass. The effective theory result is exact in the $x\to 1$ limit.  This contribution dominates the $gg \to A$ cross-section after convolution with the parton densities. It was indeed suggested that an effective theory where the bottom quark has been also integrated out could be, a posteriori, a reasonable approximation~\cite{Harlander:2003xy}.
Therefore, we argue that a matching procedure like the one suggested for the top quark, should capture most of the contribution. However, we do expect the uncertainty to be larger as uncontrolled, mass-dependent terms at moderated $x$ play a more relevant role. 
In order to asses the impact of finite fermion mass effect  we compute $K$-factors, defined as the ratio of the matched inclusive cross-section to the effective theory result, for which top and bottom are both infinitely massive. For $\tan \beta = 10$, we find:
\bea
K^{{\rm tb }}&=& 0.971, \quad m_A= 100\, {\rm GeV}, \nonumber \\
K^{{\rm tb }}&=& 0.964, \quad m_A= 200\, {\rm GeV} ,\nonumber \\
K^{{\rm tb }} &=&0.955, \quad m_A= 500\, {\rm GeV}.
\eea
The $K$-factors show that, as expected, finite mass effects are more pronounced when we include the bottom in the loop. They are of the order of a few percent at low values of the pseudoscalar mass, reaching $5 \%$ for $m_A=500$~GeV. However, we should stress that the matching procedure and the use of the effective theory at moderate $x$ are less under control than in the small $\tan \beta$ case, where we only have to deal with the top quark. For this reason these $K$-factor should be taken as an estimate.

\section{Conclusions and Outlook}
In this letter we have computed the leading small $x$ singularities of the cross-section for pseudoscalar Higgs production in gluon gluon fusion to all orders in the strong coupling constant. We have also determined the small $x$ behaviour of the Higgs rapidity distribution to the same accuracy. 
We have included the contributions of top and bottom quarks, together with their interference, showing that the cross-section only exhibits single high-energy logarithms, as opposed to the calculation in the effective theory where the quarks are integrated out.
We have explicitly evaluated the coefficients of the leading logarithms up to NNLO. 

In the spirit of Ref.~\cite{SimoneHiggs,SimoneHiggsProc}, we have used our result to construct an approximation to the NNLO cross-section, by matching the effective theory result at large $x$ to our asymptotic small $x$ behaviour. In this way we can estimate the bulk of finite fermion mass corrections. At small $\tan \beta$, where the top quark contribution dominates, we have found that they are small, reaching $1$-$2 \%$ only for $m_A \ge 500$~GeV. At large $\tan \beta$ the bottom quark contribution cannot be neglected. Including it in a similar way as we have done for the top, we have found an effect of a few percent.
In order to refine this estimate, it  would be important to match the small $x$ behaviour to the asymptotic expansion in inverse powers of the fermion mass, of which the effective theory can be seen as the zeroth order term. 
\newpage
{\bf Acknowledgments} 
We thank Stefano Forte for encouragement, useful discussions and a critical reading of the manuscript.
We also acknowledge fruitful exchanges with Matthias Steinhauser, Mikhail Rogal and Alexey Pak.  
 F.C. would like to thank the Fermilab Theory Group for their hospitality during the last stages of this work.  F.C. is supported in part by the European network HEPTOOLS under contract MRTN-CT-2006-035505, by an Italian PRIN-2008 grant and by a US-Italy Fulbright Visiting Student Researcher Fellowship. The work of S.M. is supported by UK's STFC.

\end{document}